%% file: PICOAPC.tex
\documentclass[12pt]{article}
\usepackage{amssymb,amsmath,times}
\usepackage[T1]{fontenc}
\usepackage{mathptmx}
\usepackage[dvipsnames]{xcolor}
\usepackage{graphicx}
\usepackage{fancyhdr}
\usepackage{multirow}
\usepackage[numbers,sort&compress]{natbib}
\usepackage{color}
\usepackage[nolist]{acronym}
\usepackage[plainpages=false,pdfpagelabels,colorlinks=true]{hyperref}
\usepackage[latin1]{inputenc}
\usepackage{grffile}
\usepackage{multirow}
\usepackage{adjustbox}
\usepackage{subfiles}
\usepackage[small,compact]{titlesec}
\usepackage[font=small]{caption}
\usepackage{multirow}
\usepackage{afterpage}
\usepackage{multicol}
\usepackage[normalem]{ulem}
\usepackage{wrapfig}
\usepackage{pdfpages}
\usepackage{bm}

\usepackage{cleveref}
\crefformat{footnote}{#2\footnotemark[#1]#3}

\usepackage{chngcntr}
\counterwithin{figure}{section}
\counterwithin{table}{section}

\usepackage{enumitem}
\setitemize{noitemsep,topsep=0pt,parsep=0pt,partopsep=0pt}
\setenumerate{noitemsep,topsep=0pt,parsep=0pt,partopsep=0pt}

\usepackage[font={small},margin=0em]{caption}

\usepackage[small,compact]{titlesec}
\titleformat{\section}
  {\normalfont\sffamily\Large\bfseries\color{MidnightBlue}\filcenter}
  {\thesection}{1em}{}
\titleformat{\subsection}
  {\normalfont\sffamily\large\bfseries\color{MidnightBlue}}
  {\thesubsection}{1em}{}
\titleformat{\subsubsection}
  {\normalfont\sffamily\bfseries\itshape\color{MidnightBlue}}
  {\thesubsubsection}{0.5em}{}
\titleformat{\paragraph}[runin]
  {\normalfont\sffamily\bfseries\color{MidnightBlue}}
  {\theparagraph}{}{$\bullet$\hskip 0.5em}
\titlespacing*{\subsection}{0pt}{6pt}{2pt}
\titlespacing*{\subsubsection}{0pt}{4pt}{2pt}

\fancypagestyle{cost}{\fancyhf{}\fancyfoot[C]{\thepage\\\textit{\sf\color{gray}The cost information contained in this document is of a budgetary and planning nature and is intended for informational purposes only.}}}

\newlength{\pagewidthA}
\newlength{\pageheightA}
\newlength{\pagewidthB}
\newlength{\pageheightB}
\newlength{\stockwidth}
\newlength{\stockheight}

\usepackage{geometry}
\newcommand{\generatePageLayouts}{%
  \newgeometry{layoutwidth=\pagewidthA,layoutheight=\pageheightA, left=1in,right=1in,top=1in,bottom=1in}
  \savegeometry{LayoutPageA}

  \newgeometry{layoutwidth=\pagewidthB,layoutheight=\pageheightB, left=1in,right=1in,top=.75in,bottom=1in}
  \savegeometry{LayoutPageB}
}

\newcommand{\switchToLayoutPageA}{%
  \pdfpagewidth=\pagewidthA \pdfpageheight=\pageheightA % for PDF output
  \paperwidth=\pagewidthA \paperheight=\pageheightA     % for TikZ
  \stockwidth=\pagewidthA \stockheight=\pageheightA % hyperref (memoir)?!
  \loadgeometry{LayoutPageA} % note; \loadgeometry may reset paperwidth/h!
}

\newcommand{\captiontext}{\small \setlength{\baselineskip}{0.90\baselineskip}}

\input{PICOReport_defs.tex}

\input{Planck.tex}

\begin{document}

% generate page layouts first based on layoutwidth as page size;
  % don't switch actual page sizes yet:
  \generatePageLayouts{}

\setlength{\baselineskip}{0.96\baselineskip} %% measured, 5.0 lines/inch.  Can go to 0.96\baselineskip
\setlength{\parskip}{1.\parskip}
 \switchToLayoutPageA{}

%\tableofcontents

\pagenumbering{gobble}

\parindent = 15pt

\pagenumbering{roman}
\setcounter{page}{1}

\input coverpage_apc.tex
%
{ \begin{centering}
\small
{***}\\
\smallskip
{This research was funded by a NASA grant NNX17AK52G to the University of Minnesota / Twin Cities, by the Jet Propulsion Laboratory, California Institute of Technology, under a contract with the National Aeronautics and Space Administration, and by Lockheed Martin Corporation. \\ Substantial contributions to the development of PICO were volunteered by scientists at many institutions world-wide.  They are very gratefully acknowledged.} \\
\medskip

{The information presented about the PICO mission concept is pre-decisional and is provided for planning and discussion purposes only.}\\
\medskip

{The cost information contained in this document is of a budgetary and planning nature and is intended for informational purposes only.  It does not constitute a commitment on the part of JPL and/or Caltech.}\\
%\bigskip

\end{centering} }

%\newpage

%{
%             % KY: making table of contents links black instead of red.
%\hypersetup{linkcolor=black}
%\tableofcontents
%}
\newpage
\pagenumbering{arabic}
\setcounter{page}{1}
\setcounter{figure}{0}

\section{Executive Summary} % (2 pg, Hanany)}

\vspace{-0.04in}

\input executive2_apc.tex

\vspace{-0.07in}

\section{Key Science Goals and Objectives}
\label{sec:science}

\vspace{-0.06in}

\input fundamentalsci_apc.tex

\vspace{-0.06in}

\subsection{Cosmic Structure Formation and Evolution} % (4 pgs)
\label{sec:extragalacticsci}

\vspace{-0.03in}

\input extragalacticsci2_apc.tex
\vspace{-0.06in}

\subsection{Testing $\Lambda$CDM} % (3 pgs)
\label{sec:testinglcdm}

\vspace{-0.08in}

\input testlcdm-apc.tex

% ------------

\vspace{-0.06in}

\subsection{Galactic Structure and Star Formation} % (3 pgs)
\label{sec:galacticsci}

%\vspace{-0.06in}

\input galacticsci-apc.tex

% ------------

\vspace{-0.06in}

\subsection{Legacy Surveys} % (2 pgs)
\label{sec:legacy}

%\input{stm2.tex}
\input Legacy-apc.tex

% ------------
\vspace{-0.06in}

\subsection{Foregrounds and Systematics}%  (4 pages)}
\label{sec:signal_separation}

\input foregrounds-systematics-apc.tex

\vspace{-0.06in}

\input technical-apc.tex

\vspace{-0.06in}

\section{Technology Drivers}
\label{sec:techdrivers}

\vspace{-0.06in}

\input tech-drivers-apc.tex

\vspace{-0.06in}

\input organization-apc.tex

\vspace{-0.06in}

\input schedule-and-cost-apc.tex

\newpage
\def\bibfont{\footnotesize}
\setlength{\bibsep}{1pt}
\bibliographystyle{IEEEtranN}
\bibliography{mybib,pico-report}

\input{acronym}

\end{document}

%% file: PICOReport_defs.tex
\def\bei{\begin{itemize}}
\def\eei{\end{itemize}}

\def\Neff{{N_{\rm eff}}}

\def\planck{{\it Planck}}
\def\jwst{{\it JWST}}
\def\erosita{{\it eROSITA}}
\def\euclid{{\it Euclid}}
\def\gaia{{\it Gaia}}

\def\wmap{{\it WMAP}}
\def\cobe{{\it COBE}}

\newcommand{\degree}{\ensuremath{^\circ}}

\def\ruo2{RuO$_{2}$}

\def\mathrelfun#1#2{\lower3.6pt\vbox{\baselineskip0pt\lineskip.9pt
  \ialign{$\mathsurround=0pt#1\hfil##\hfil$\crcr#2\crcr\sim\crcr}}}

\def\simgt{\mathrel{\mathpalette\mathrelfun >}}

\long\def\comment#1{}

\newbox\tablebox    \newdimen\tablewidth
\def\leaderfil{\leaders\hbox to 5pt{\hss.\hss}\hfil}

\def\tablenote#1 #2\par{\begingroup \parindent=0.8em
    \abovedisplayshortskip=0pt\belowdisplayshortskip=0pt
    \noindent
    $$\hss\vbox{\hsize\tablewidth \hangindent=\parindent \hangafter=1 \noindent
    \hbox to \parindent{$^#1$\hss}\strut#2\strut\par}\hss$$
    \endgroup}
\def\doubleline{\vskip 3pt\hrule \vskip 1.5pt \hrule \vskip 5pt}
\newcommand{\wisk}[1]{{\ifmmode{#1}\else{$#1$}\fi}}

%% file: Planck.tex
\newbox\tablebox    \newdimen\tablewidth
\def\leaderfil{\leaders\hbox to 5pt{\hss.\hss}\hfil}
\def\endPlancktable{\tablewidth=\columnwidth 
    $$\hss\copy\tablebox\hss$$
    \vskip-\lastskip\vskip -2pt}

\def\tablenote#1 #2\par{\begingroup \parindent=0.8em
    \abovedisplayshortskip=0pt\belowdisplayshortskip=0pt
    \noindent
    $$\hss\vbox{\hsize\tablewidth \hangindent=\parindent \hangafter=1 \noindent
    \hbox to \parindent{$^#1$\hss}\strut#2\strut\par}\hss$$
    \endgroup}
\def\doubleline{\vskip 3pt\hrule \vskip 1.5pt \hrule \vskip 5pt}

%% file: coverpage_apc.tex
%\documentclass[PICOAPC.tex]{subfiles}

%\begin{document}
%
% author list table.
% 4 columns, authors + affiliations
%
\LARGE{ \centerline{\bf{PICO: Probe of Inflation and Cosmic Origins}}}
%
%\vskip -16pt
%\Large{ \centerline{Report from a Probe-Scale Mission Study}}
%\Large{ \centerline{July, 2019 }}

\parindent = 0pt

\vspace{4pt}
\large{Thematic Area: Space Based Projects - Medium}

\vspace{2pt}
\parindent = 0pt
%\normalsize{
\normalsize{ Lead Author: Shaul Hanany (hanany@umn.edu)\,$^{1}$ } 

\vspace{-0.06in} 

%\normalsize{
%\small{
%Steering Committee: Charles Bennett\,$^{2}$, Scott Dodelson\,$^{3}$, Lyman Page\,$^{4}$ } \\
%\vspace{-6pt} \\
%\normalsize{
%\small{
%\raggedright
% Executive Committee:
% James~Bartlett\,$^{5,6}$,
% Nick~Battaglia\,$^{7}$,
% Jamie~Bock\,$^{8,6}$,
% Julian~Borrill\,$^{9,10}$,
% David~Chuss\,$^{11}$,
% Brendan~P.~Crill\,$^{6}$,
% Jacques~Delabrouille\,$^{5,12}$,
% Mark~Devlin\,$^{13}$,
% Laura~Fissel\,$^{14}$,
% Raphael~Flauger\,$^{15}$,
% Dan~Green\,$^{16}$,
% J.~Colin~Hill\,$^{17,18}$,
% Johannes~Hubmayr\,$^{19}$,
% William~Jones\,$^{4}$,
% Lloyd~Knox\,$^{20}$,
% Al~Kogut\,$^{21}$,
% Charles~Lawrence\,$^{6}$,
% Jeff~McMahon\,$^{22}$,
% Tim~Pearson\,$^{8}$,
% Clem~Pryke\,$^{1}$,
% Marcel~Schmittfull\,$^{18}$,
% Amy~Trangsrud\,$^{6}$,
% Alexander~van~Engelen\,$^{23}$
% \\
% }
\label{authorlist}

\vspace{6pt}

\large  {\centerline {Authors}}
%\vskip 0pt
%
\setlength{\columnsep}{7pt}
\footnotesize {
% change to left justified
\begin{multicols}{4}
Marcelo Alvarez $^{2,3}$                 \\
Emmanuel Artis $^{4}$                  \\
%Peter Ashton $^{9}$                   \\
Peter Ashton $^{2,3,5}$                    \\
%Peter Ashton $^{24}$                   \\
Jonathan Aumont $^{6}$                 \\
Ragnhild Aurlien $^{7}$                \\
Ranajoy Banerji $^{7}$                 \\
R. Belen Barreiro $^{8}$               \\
James G. Bartlett $^{9,10}$               \\
%James G. Bartlett $^{6}$              \\
Soumen Basak $^{11}$                    \\
Nick Battaglia $^{12}$                  \\
Jamie Bock $^{10,13}$                      \\
%Jamie Bock   !!Caltech then JPL!! $^{6}$   \\
Kimberly K. Boddy $^{14}$               \\
Matteo Bonato $^{15}$                   \\
Julian Borrill $^{2,16}$                  \\
%Julian Borrill $^{10}$                 \\
Fran\c{c}ois Bouchet $^{17}$            \\
Fran\c{c}ois Boulanger $^{18}$          \\
Blakesley Burkhart $^{19}$              \\
Jens Chluba $^{20}$                     \\
David Chuss $^{21}$                     \\
Susan E. Clark $^{22,23}$                  \\
Joelle Cooperrider $^{10}$              \\
Brendan P. Crill $^{10}$                \\
Gianfranco De Zotti $^{24}$             \\
Jacques Delabrouille $^{4,9}$            \\
%Jacques Delabrouille $^{12}$           \\
%Mark Devlin $^{13}$   \\
Eleonora Di Valentino $^{25}$           \\
Joy Didier $^{26}$                      \\
Olivier Dor\'e $^{10,13}$                  \\
%Olivier Dor\'e    !!! JPL then Caltech!!! $^{8}$   \\
Hans K. Eriksen $^{7}$                 \\
Josquin Errard $^{9}$                  \\
Tom Essinger-Hileman $^{27}$            \\
Stephen Feeney $^{28}$                  \\
Jeffrey Filippini $^{29}$               \\
Laura Fissel $^{30}$                    \\
Raphael Flauger $^{31}$                 \\
Unni Fuskeland $^{7}$                  \\
%Vera Gluscevic $^{39}$                  \\
Vera Gluscevic $^{26}$                  \\
Krzysztof M. Gorski $^{10}$             \\
Dan Green $^{3}$                       \\
Shaul Hanany $^{1}$                    \\
Brandon Hensley $^{32}$                 \\
Diego Herranz $^{8}$                   \\
J. Colin Hill $^{22,28}$                   \\
%J. Colin Hill $^{18}$                  \\
Eric Hivon $^{17}$                      \\
Ren\'{e}e  Hlo\v{z}ek $^{33}$           \\
Johannes Hubmayr $^{34}$                \\
Bradley R. Johnson $^{35}$              \\
William Jones $^{32}$                   \\
Terry Jones $^{1}$                     \\
Lloyd Knox $^{36}$                      \\
Al Kogut $^{27}$                        \\
Marcos L\'{o}pez-Caniego $^{37}$        \\
Charles Lawrence $^{10}$                \\
Alex Lazarian $^{38}$                   \\
Zack Li $^{32}$                         \\
Mathew Madhavacheril $^{32}$            \\
%Jeff McMahon $^{22}$   \\
Jean-Baptiste Melin $^{4}$             \\
Joel Meyers $^{39}$                     \\
Calum Murray $^{9}$                    \\
Mattia Negrello $^{40}$                 \\
Giles Novak $^{41}$                     \\
Roger O'Brient $^{10,13}$                  \\
%Roger O'Brient      !!! JPL then Caltech!!! $^{8}$   \\
Christopher Paine $^{10}$               \\
Tim Pearson $^{13}$                     \\
Levon Pogosian $^{42}$                  \\
Clem Pryke $^{1}$                      \\
%Giuseppe Puglisi $^{48}$               \\
Giuseppe Puglisi $^{43,44}$                \\
Mathieu Remazeilles $^{20}$             \\
Graca Rocha $^{10,13}$                     \\
%Graca Rocha $^{8}$                    \\
Marcel Schmittfull $^{22}$              \\
Douglas Scott $^{45}$                   \\
Peter Shirron $^{27}$                   \\
Ian Stephens $^{46}$                    \\
Brian Sutin $^{10}$                     \\
Maurizio Tomasi $^{47}$                 \\
Amy Trangsrud $^{10}$                   \\
Alexander van Engelen $^{48}$           \\
Flavien Vansyngel $^{49}$               \\
Ingunn K. Wehus $^{7}$                 \\
Qi Wen $^{1}$                          \\
Siyao Xu $^{38}$                        \\
Karl Young $^{1}$                      \\
Andrea Zonca $^{50}$
\end{multicols}
}

\vskip -7pt

\large { \centerline {Endorsers}}

%\vskip 0pt
%
\setlength{\columnsep}{7pt}
\footnotesize {
%\begin{minipage}{1.2\textwidth}
\begin{multicols}{4}
Maximilian Abitbol              \\
Zeeshan Ahmed                   \\
David Alonso                    \\
Mustafa A. Amin                 \\
Adam Anderson                   \\
James Annis                     \\
Jason Austermann                \\
Carlo Baccigalupi               \\
Darcy Barron                    \\
Ritoban Basu Thakur             \\
Elia Battistelli                \\
Daniel Baumann                  \\
Karim Benabed                   \\
Bradford Benson                 \\
Paolo de Bernardis              \\
Marco Bersanelli                \\
Federico Bianchini              \\
Daniel Bilbao-Ahedo             \\
Colin Bischoff                  \\
Sebastian Bocquet               \\
J. Richard Bond                 \\
Jeff Booth                      \\
Sean Bryan                      \\
Carlo Burigana                  \\
Giovanni Cabass                 \\
Robert Caldwell                 \\
John Carlstrom                  \\
Xingang Chen                    \\
Francis-Yan Cyr-Racine          \\
Paolo de Bernardis              \\
Tijmen de Haan                  \\
C. Darren Dowell                \\
Cora Dvorkin                    \\
Chang Feng                      \\
Ivan Soares Ferreira            \\
Aurelien Fraisse                \\
Andrei V. Frolov                \\
Nicholas Galitzki               \\
Silvia Galli                    \\
Ken Ganga                       \\
Tuhin Ghosh                     \\
Sunil Golwala                   \\
Riccardo Gualtieri              \\
Jon E. Gudmundsson              \\
Nikhel Gupta                    \\
Nils Halverson                  \\
Kyle Helson                     \\
Sophie Henrot-Versill\'e        \\
Thiem Hoang                     \\
Kevin M. Huffenberger           \\
Kent Irwin                      \\
Reijo Keskitalo                 \\
Rishi Khatri                    \\
Chang-Goo Kim                   \\
Theodore Kisner                 \\
Arthur Kosowsky                 \\
Ely Kovetz                      \\
Kerstin Kunze                   \\
Guilaine Lagache                \\
Daniel Lenz                     \\
Fran\c{c}ois Levrier            \\
Marilena Loverde                \\
Philip Lubin                    \\
Juan Macias-Perez               \\
Nazzareno Mandolesi             \\
Enrique Mart\'{i}nez-Gonz\'{a}lez   \\
Carlos Martins                  \\
Silvia Masi                     \\
Tomotake Matsumura              \\
Darragh McCarthy                \\
P. Daniel Meerburg              \\
Alessandro Melchiorri           \\
Marius Millea                   \\
Amber Miller                    \\
Joseph Mohr                     \\
Lorenzo Moncelsi                \\
Pavel Motloch                   \\
Tony Mroczkowski                \\
Suvodip Mukherjee               \\
Johanna Nagy                    \\
Pavel Naselsky                  \\
Federico Nati                   \\
Paolo Natoli                    \\
Michael Niemack                 \\
Elena Orlando                   \\
Bruce Partridge                 \\
Marco Peloso                    \\
Francesco Piacentini            \\
Michel Piat                     \\
Elena Pierpaoli   \\
Giampaolo Pisano                \\
Nicolas Ponthieu                \\
Giuseppe Puglisi                \\
Benjamin Racine                 \\
Christian Reichardt             \\
Christophe Ringeval             \\
Karwan Rostem                   \\
Anirban Roy                     \\
Jose Alberto Rubino-Martin      \\
Matarrese Sabino                \\
Maria Salatino                  \\
Benjamin Saliwanchik            \\
Neelima Sehgal                  \\
Sarah Shandera                  \\
Erik Shirokoff                  \\
An\v{z}e Slosar                 \\
Tarun Souradeep                 \\
Suzanne Staggs                  \\
George Stein                    \\
Radek Stompor                   \\
Rashid Sunyaev                  \\
Aritoki Suzuki                  \\
Eric Switzer                    \\
Andrea Tartari                  \\
Grant Teply                     \\
Peter Timbie                    \\
Matthieu Tristram               \\
Caterina Umilt\`{a}             \\
Rien van de Weygaert            \\
Vincent Vennin                  \\
Licia Verde                     \\
Patricio Vielva                 \\
Abigail Vieregg                 \\
Jan Vrtilek                     \\
Benjamin Wallisch               \\
Benjamin Wandelt                \\
Gensheng Wang                   \\
Scott Watson                    \\
Edward J. Wollack               \\
Zhilei Xu                       \\
Siavash Yasini
%Martin White    \\
\end{multicols}
%\end{minipage}
}

\vskip -7pt

\Large  {\centerline {Affiliations}}
\vspace{-6pt}
\begin{multicols}{2}
\raggedright
\scriptsize {
%\footnotesize {
1. University of Minnesota - Twin Cities.  \\
2. Lawrence Berkeley National Laboratory.  \\
3. University of California, Berkeley.  \\
4. IRFU, CEA, Universit\'e Paris-Saclay, France.  \\
5. Kavli Institute for the Physics and Mathematics of the Universe (WPI).  \\
6. IRAP, Universit\'e de Toulouse, France.  \\
7. University of Oslo, Norway.  \\
8. Instituto de F\'isica de Cantabria (CSIC-Universidad de Cantabria), Spain.  \\
9. APC, Univ Paris Diderot, CNRS/IN2P3, CEA/lrfu, Obs de Paris, Sorbonne Paris Cit\'e, France.  \\
10. Jet Propulsion Laboratory, California Institute of Technology.  \\
11. School of Physics, Indian Institute of Science Education and Research Thiruvananthapuram, India.  \\
12. Cornell University.  \\
13. California Institute of Technology.  \\
14. Johns Hopkins University.  \\
15. INAF-Istituto di Radioastronomia and Italian ALMA Regional Centre, Italy.  \\
16. Space Sciences Laboratory, University of California, Berkeley.  \\
17. Institut d'Astrophysique de Paris, CNRS and Sorbonne Universit\'e, France.  \\
18. Ecole Normale Superieure, Paris, France.  \\
19. Rutgers University.  \\
20. JBCA, University of Manchester.  \\
21. Villanova University.  \\
22. Institute for Advanced Study, Princeton.  \\
23. Hubble Fellow          \\
24. INAF-Osservatorio Astronomico di Padova, Italy.  \\
25. University of Manchester.  \\
26. University of Southern California.  \\
27. NASA Goddard Space Flight Center.  \\
28. Center for Computational Astrophysics, Flatiron Institute.  \\
29. University of Illinois, Urbana-Champaign.  \\
30. National Radio Astronomy Observatory.  \\
31. University of California, San Diego.  \\
32. Princeton University.  \\
33. Department of Astronomy \& Astrophysics and Dunlap Institute, University of Toronto, Canada.  \\
34. National Institute of Standards and Technology.  \\
35. Columbia University.  \\
36. University of California, Davis.  \\
37. European Space Astronomy Centre.  \\
38. University of Wisconsin - Madison.  \\
39. Southern Methodist University.  \\
40. Cardiff University School of Physics and Astronomy.  \\
41. Northwestern University.  \\
42. Simon Fraser University.  \\
43. Stanford University.  \\
44. Kavli Institute for Particle Astrophysics and Cosmology.  \\
45. University of British Columbia, Canada.  \\
46. Harvard-Smithsonian Center for Astrophysics.  \\
47. Universit\`a degli studi di Milano.  \\
48. Canadian Institute for Theoretical Astrophysics, University of Toronto, Canada.  \\
49. Institut d'Astrophysique Spatiale, CNRS, Univ. Paris-Sud, Universit\'e Paris-Saclay, France.  \\
50. San Diego Supercomputer Center, University of California, San Diego.

%3. Carnegie Melon University.  \\
%13. University of Pennsylvania.  \\
%22. University of Michigan.  \\

}
\end{multicols}

\vskip -7pt

\normalsize
%\end{document}

%% file: executive2_apc.tex
%\documentclass[PICOAPC.tex]{subfiles}

%\begin{document}

The Probe of Inflation and Cosmic Origins (PICO) is a proposed probe-scale space mission\footnote{PICO was selected by NASA to conduct a probe mission study. The full report is available~\citep{pico_report,picoweb,nasa_decadalweb}.} consisting of an imaging polarimeter that will scan the sky for 5 years in 21 frequency bands from 21 to 799~GHz. It will produce full-sky surveys of intensity and polarization with a final combined-map noise level equivalent to 3300 \planck\ missions for the baseline required specifications, and according to our current best-estimate would perform as 6400 \planck\ missions. With these capabilities, only available in space and unmatched by any other existing or proposed platform: \\
$\bullet$ PICO could determine the energy scale of inflation and give a first, direct probe of quantum gravity by searching for the signal that arises from gravitational waves sourced by inflation and parameterized by the tensor-to-scalar ratio $r$. The PICO requirement is to detect $r =5\times10^{-4} \, (5\sigma)$, a level that is 100 times lower than current upper limits, and 5 times lower than limits forecast by any planned experiment.  If the signal is not detected, PICO is the only instrument that can exclude at $5 \sigma$ models for which the characteristic scale in the potential is the Planck scale, a key threshold in inflation physics. \\ %\comred{cite swp} \\
$\bullet$ The mission will measure the minimum expected sum of the neutrino masses with $4\sigma$ confidence, rising to $7\sigma$ if the sum is near 0.1~eV. \\ 
%Reaching the $4\sigma$ level can only be achieved with an instrument that can measure the polarization of the CMB on the largest angular scales {\it and} a measurement best done from space, which gives access to the full sky, and with a broad band of frequencies to remove foreground contaminants.  
%PICO will give two additional independent and equally competitive constraints on the sum of neutrino masses. \\
$\bullet$ The measurements will either detect or strongly constrain deviations from the standard model of particle physics by counting the number of light particle species $N_{\rm eff}$ in the early universe with $\Delta N_{\rm eff} < 0.06 \, (2\sigma)$.  \\
$\bullet$ PICO will elucidate the processes affecting the evolution of cosmic structures by measuring the optical depth to reionization $\tau$ with an error $\sigma(\tau) = 0.002$, limited only by the number of spatial modes available in the largest angular scale \ac{CMB} polarization. \\
$\bullet$ The data will give a full sky map of the projected gravitational potential due to all structures in the Universe with the highest \ac{SNR} relative to any foreseeable experiment, and it will give a catalog of 150,000 clusters extending to their earliest formation redshift. Each of these datasets will be used in combination with other data to constrain the evolution of the amplitude of linear fluctuations $\sigma_{8}(z)$ with sub-percent accuracy and thus constrain dark energy and modified gravity models.  \\
$\bullet$ PICO will determine the cosmological paradigm of the 2030's by reducing the allowed volume of uncertainty in an 11-dimensional $ \Lambda$CDM parameter space by a factor of nearly a billion relative to current \planck\ constraints. % on only six parameters. 
Such exquisite scrutiny will either give strong validation of the model or require yet-to-be discovered revisions. \\
$\bullet$ With 86,000,000 independent polarization measurements across the Milky Way, 2,900 times more than \planck \ had, PICO's data will be used to resolve long-standing questions about our Galaxy, including the 
composition, temperature, and emissivities of Galactic dust, and the relative roles of gas turbulence and magnetic fields in the dynamics of the Galaxy and in the observed low star-formation efficiency. \\
$\bullet$ The data will constrain generic models of dark matter, enable a search for primordial magnetic fields with sufficient sensitivity to rule them out as the sole source for the largest observed Galactic magnetic fields, constrain string-theory-motivated axions, 
%by improving by a factor of 300 constraints on polarization rotation arising from early Universe fields;  
and give precise tracing of the evolution with $z$ of thermal pressure in the Universe. \\ 
%through correlations of the thermal Sunyaev--Zeldovich effect with LSST's gold sample of galaxies, which will exceed a signal-to-noise ratio of 1000. \\
$\bullet$ PICO's deep, full-sky legacy maps will constrain the early phases of galaxy and cluster evolution,  
perform a census of cold dust in thousands of low-$z$ galaxies, make cosmic infrared background maps 
due to dusty star-forming galaxies, and map magnetic fields in 70 nearby galaxies. 

With its broad frequency coverage, PICO is better equipped than any other current or planned instrument to separate the detected signals into their original sources of emission.  This capability is important for many of the science goals, and may be key for unveiling the faintest of signals, the telltale signature of inflation, which is already known to be dominated by Galactic foregrounds. 
PICO's large multiplicity of independent maps and sky surveys, and its stable thermal environment will give control of systematic uncertainties unmatched by any other platform. 
%It will conduct observations from L2, and execute ten redundant,  full-sky surveys, each complete within 6 months. 
Mission operations are simple: PICO has a single instrument that surveys the sky with a continuously repetitive pattern.  
The required technologies have either already been proven by past missions, or are extensions of technologies now being used by sub-orbital experiments.  

The science PICO will deliver addresses some of the most fundamental quests of human knowledge. The advances it will make will enrich many areas of astrophysics, and will form the basis for the cosmological paradigm of the 2030's and beyond. 
Progress in sub-orbital CMB science requires a scale-up of investment with some proposed costs nearing those of a Probe-class space mission. 
%Progress in CMB science requires a scale-up of investment. 
PICO is the most cost-effective way to make progress as it has no competitor in terms of raw sensitivity, and it is the only single-platform instrument with the combination of angular resolution, frequency bands, and control of systematic effects that can deliver the compelling, timely, and broad science. 

%PICO is the most cost-effective way to achieve this scale-up. It has no competitor in terms of raw sensitivity, and it is the only single-platform instrument with the combination of angular resolution, frequency bands, and control of systematic effects that can deliver the compelling, timely, and broad science. 

% Progress in ground-based CMB science requires a scale-up of investment with proposed costs nearing those of a Probe-class space mission. PICO is a more cost-effective means to make progress as it has no competitor in terms of raw sensitivity, and it is the only single-platform instrument with the combination of angular resolution, frequency bands, and control of systematic effects that can deliver the compelling, timely, and broad science. 

%\end{document}

%% file: fundamentalsci_apc.tex
%\documentclass[PICOAPC.tex]{subfiles}

%\begin{document}

\subsection{Gravitational Waves and Inflation}
\label{sec:inflation}

\begin{figure}[!thb]
\vspace{-.1in}
\hspace{-0.13in}
\parbox{4.4in}{\centerline{
\includegraphics[width=4.5in]{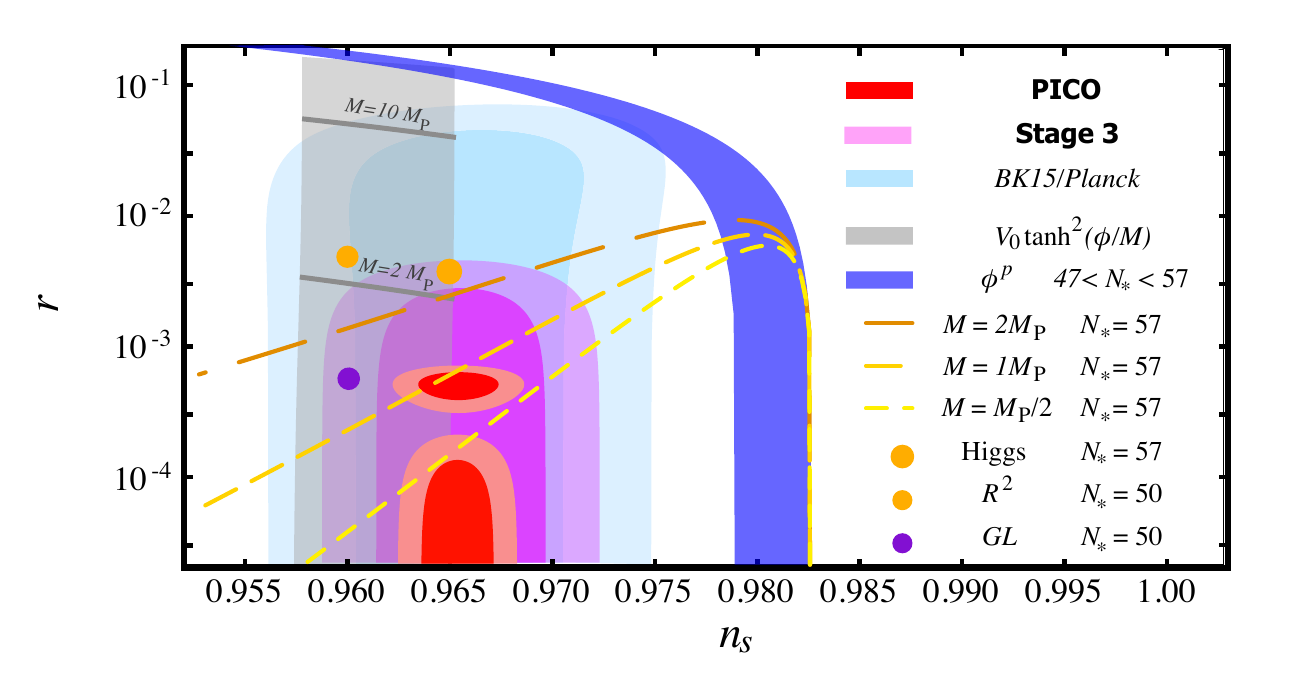} } }
\parbox{2.1in}{
\caption{\captiontext  PICO will conclusively rule out all inflation models for which the characteristic scale in the potential is $M_{\rm P}$ or higher (dashed lines), or will detect $r=0.0005$ at $5\, \sigma$ (red 68\% and 95\% limits and uncertainty ellipses). Current values of $\sigma_{r}$ are a factor of 100 higher (cyan). 
The loci of classes of models and of specific ones are shown with dots, solids, and shades. }
\label{fig:nsr}}
\vspace{-0.16in}
\end{figure}

Measurements of the \ac{CMB} $BB$ angular power spectrum are the only foreseeable way to detect  inflationary gravitational waves. The strength of the signal, quantified by the tensor-to-scalar ratio $r$, is a direct measure of the expansion rate of the Universe during inflation; together with the Friedmann equation, it reveals the energy scale of inflation. PICO will detect primordial gravitational waves at $5\,\sigma$ significance if inflation occurred at an energy scale of at least $5\times 10^{15}\,\rm{GeV}$, or equivalently if $r= 5\times 10^{-4}$.  In a widely endorsed community white paper setting targets for measurements of inflationary gravitational waves in the next decade, \citet{Shandera_etal} quote two theoretically motivated $r$ rejection targets: (1) $r < 0.01$, and (2) $r < 0.001$. The second threshold is motivated by the goal of rejecting all inflationary models that naturally explain the observed value of the spectral index $n_{\rm s}$ and having a characteristic scale in the potential that is larger than the Planck scale. Such models are shown as dashed lines in Figure~\ref{fig:nsr}.  They write "If these thresholds are passed without a detection, most textbook models of inflation will be ruled out; and ... the data would then force a significant change in our understanding of the primordial Universe." PICO is the only next-decade experiment with the raw sensitivity to reject both targets at high confidence. 
%; see Figure~\ref{fig:nsr}. It is the only next-decade experiment that can detect inflationary models that have $r \geq 5\times 10^{-4}$ at high confidence. 

%PICO will be able to detect gravitational waves from inflation at $5\sigma$ greater if $r = 0.0005$ or will exclude all single-field models that naturally explain the observed value of the spectral index with a characteristic scale in the potential of $M_{\rm p}$ or higher (dashed lines). The red contours show 1 and $2\sigma$ limits and uncertainty ellipses. Current values of $\sigma(r)$ are a factor of 100 higher (cyan). Predictions for the classes of models that naturally explain the spectral index and sample individual models are shown with dots, solids, and shades.

Uncertainty in the characterization of Galactic foregrounds already limits our ability to constrain $r$. These foregrounds 
are anticipated to be nearly 1000 times stronger than next-decade-targeted inflationary $B$-mode signals at low $\ell \simeq 8$ multipoles. `Lensing' $B$-modes, created by gravitational lensing of $E$-modes, are an additional effective foreground for the higher $\ell \simeq 80$ multipoles. With sufficiently high resolution to remove at least 73\% of the lensing effects, and 21 frequency bands to account for foregrounds, PICO is better equipped than all other next-decade experiments to reject intervening signals. 
%robustly finding the faint inflationary signal, or in rejecting confusion due to foregrounds. 

\vspace{-0.06in}

\subsection{Fundamental Particles and Fields} %: Light Relics, Dark Matter, and Neutrinos}
\label{sec:relics_neutrinos}

%\vspace{-0.05in}

$\bullet$ {\bf Light Relics} \hspace{0.1in} The effective number of light relic particle species $\Neff$ gives information about particle species that are predicted to have existed in the early Universe in extensions of the Standard Model. Light particles beyond the three neutrino families contribute a change $\Delta \Neff$ that is a function only of the decoupling temperature of the additional species and the spin of the particle. PICO will provide a constraint $\Delta \Neff < 0.06 \, (95\%)$ and will either detect new particle species, or constrain the lowest decoupling temperature at which any spin-1 particle could have fallen out of equilibrium by a factor of 400 higher than today's constraint~\citep{green_swp}. No next-decade experiment will provide a tighter constraint.   \\ 
$\bullet$ {\bf Neutrino Mass} \hspace{0.1in} \label{neutrino_fundamental} The origin, structure, and values of the neutrino masses are among the outstanding questions about the nature of the Standard Model of particle physics.  
Cosmological measurements of $\sum m_\nu$ relate the amplitudes of the matter power spectrum and the primordial fluctuation power spectrum $A_{\rm s}$.  Both are limited by degeneracies with other parameters. PICO is the only instrument that will self consistently provide three of the four necessary measurement ingredients: $\tau$, $A_{\rm s}$, and the matter power spectrum via CMB lensing.~\citep{green_swp,dvorkin_swp}. 
In combination with DESI and \euclid~data, PICO will give $\sigma(\sum m_\nu) = 14$ meV, yielding a $4\,\sigma$ detection of the minimum sum of 58~meV. Moreover, PICO will measure  $\sum m_\nu$ in two additional ways, which will give equivalent constraints. \\
$\bullet$ {\bf Dark Matter} \hspace{0.1in} \ac{CMB} experiments are effective in constraining dark matter candidates in the lower mass range, which is not available for terrestrial direct detection experiments~\citep{Slatyer2009,Galli2009,Huetsi2009,Huetsi2011,Madhavacheril:2013cna,Green:2018pmd}. 
For a spin- and velocity-independent contact interaction between dark matter and protons, PICO will improve upon \planck 's dark matter cross-section constraints by a factor of 25 over a broad range of candidate dark matter masses. If 2\% of the total dark matter content is made of axions in the mass range $10^{-30} < m_{a} < 10^{-26}$~eV, then PICO will detect this species at between $7$ and $13\,\sigma$.  These constraints are stronger than for all other proposed next-decade CMB experiments~\citep{gluscevic_swp}. \\
$\bullet$ {\bf Primordial Magnetic Fields (PMFs)} \hspace{0.1in} PICO is the only experiment that can probe PMFs as weak as 0.1~nG ($1\,\sigma$).  Detection of PMFs would be a major discovery because it would signal new physics beyond the Standard Model of particle physics, discriminate among different theories of the early Universe, and explain the puzzling 1-10~$\mu$G fields observed in galaxies.  Or it could conclusively rule out a purely primordial (i.e., no-dynamo-driven) origin of the largest Galactic magnetic fields~\citep{Widrow:2002ud,Widrow:2011hs,Athreya:1998,Grasso:2000wj,Vachaspati:1991nm,Turner:1987bw,Ratra:1991bn,DiazGil:2007dy,Barnaby:2012tk,Long:2013tha,Durrer:2013pga}. \\
$\bullet$ {\bf Cosmic Birefringence} \hspace{0.1in}
A number of well-motivated extensions of the Standard Model involve fields with parity-violating coupling~\citep{Freese:1990rb,Frieman:1995pm,Carroll:1998zi,Kaloper:2005aj,2008PhRvL.101n1101C,Gluscevic:2010vv}. Their presence may cause cosmic birefringence -- a rotation of the polarization of an electromagnetic wave as it propagates across cosmological distances~\cite{Harari:1992ea,Carroll:1989vb,Carroll:1998zi}. PICO's constraints on cosmic birefringence are more stringent than those of any other next-decade experiment~\cite{pogosian_2019}.

%% file: extragalacticsci2_apc.tex
%\documentclass[PICOAPC.tex]{subfiles}

%\begin{document}
 
$\bullet$ {\bf The Formation of the First Luminous Sources} \hspace{0.1in} \label{sec:luminoussources}  Measurements of the optical depth to reionization $\tau$ will illuminate the nature of the first luminous sources and the exact history of the reionization epoch, both of which are key missing links in our understanding of structure formation~\citep{alvarez_swp}.  With full sky coverage, multiple frequency bands, and ample sensitivity to remove foregrounds, PICO is uniquely suited to reach cosmic-variance-limited precision, with $\sigma(\tau)=0.002$. Data from PICO's frequency bands above 400~GHz -- which have better than 2~arcmin resolution  -- will be used to provide clean maps for higher-resolution ground-based instruments that can reconstruct the patchy $\tau$ field. No other experiment can provide these data. \\
$\bullet$ {\bf Probing the Evolution of Structures via Gravitational Lensing and Cluster Counts} \hspace{0.1in} \label{sec:gravitationallensing}   
%The amplitude of linear fluctuations as a function of redshift, parameterized by $\sigma_8(z)$, is a sensitive probe of physical processes affecting growth of structures in the Universe. 
PICO will give sub-percent constraints on $\sigma_8(z)$, the amplitude of linear fluctuations as a function of redshift, through measurements of gravitational lensing of the CMB photons and independently by using cluster counts. 
PICO will have an \ac{SNR} of more than 560 for measurement of $C_{L}^{\phi \phi}$, the angular power spectrum of the projected gravitational potential $\phi$ that is lensing the photons. This is the highest of any foreseeable CMB experiment in the range $2 \leq L \lesssim 1500$. When combined with LSST data the measurement will give $\sigma_8(z) <0.5\%$ in each of six redshift bins for $z>0.5$~\cite{pico_report}.
The mission will find 150,000 galaxy clusters, and this catalog will provide $\sigma_{8}(z) < 1\%$ for each of eight bins in $0.5 < z < 2$, and a neutrino mass constraint $\sigma(\sum m_{\nu}) = 14$~meV that is independent from the one coming from $C_{L}^{\phi \phi}$. A significant fraction of the PICO-detected clusters will also be detected by \erosita, giving an exceptional catalog of multi-wavelength observations for detailed studies of cluster astrophysics. The constraints on $\sigma_{8}$ will translate to constraints on dark energy, modified gravity, baryonic feedback process, and limits on the particle content of the Universe. \\
$\bullet$ {\bf Constraining Feedback Processes through the Sunyaev--Zeldovich Effect} \hspace{0.1in} \label{sec:sz}
The thermal SZ (tSZ) effect probes the integrated electron pressure along the line-of-sight.  PICO will detect 150,000 clusters through their tSZ signature, the largest catalog of any proposed CMB experiment, including thousands of high-redshift objects that are undetectable via X-ray emission.  PICO will also provide the only full-sky, high-\ac{SNR} tSZ map of any proposed CMB experiment.  The cross-correlation of this map with the LSST gold weak-lensing sample (26 gal/arcmin$^2$ over 40\% of the sky) will be detected at \ac{SNR}=3000, yielding a precise tomographic reconstruction of the evolution of thermal pressure over cosmic time.

%% file: testlcdm-apc.tex
%\documentclass[PICOAPC.tex]{subfiles}

%\begin{document}

\begin{figure}[h]
\hspace{-0.2in}
\parbox{2.6in}{\centerline {
\includegraphics[width=2.5in]{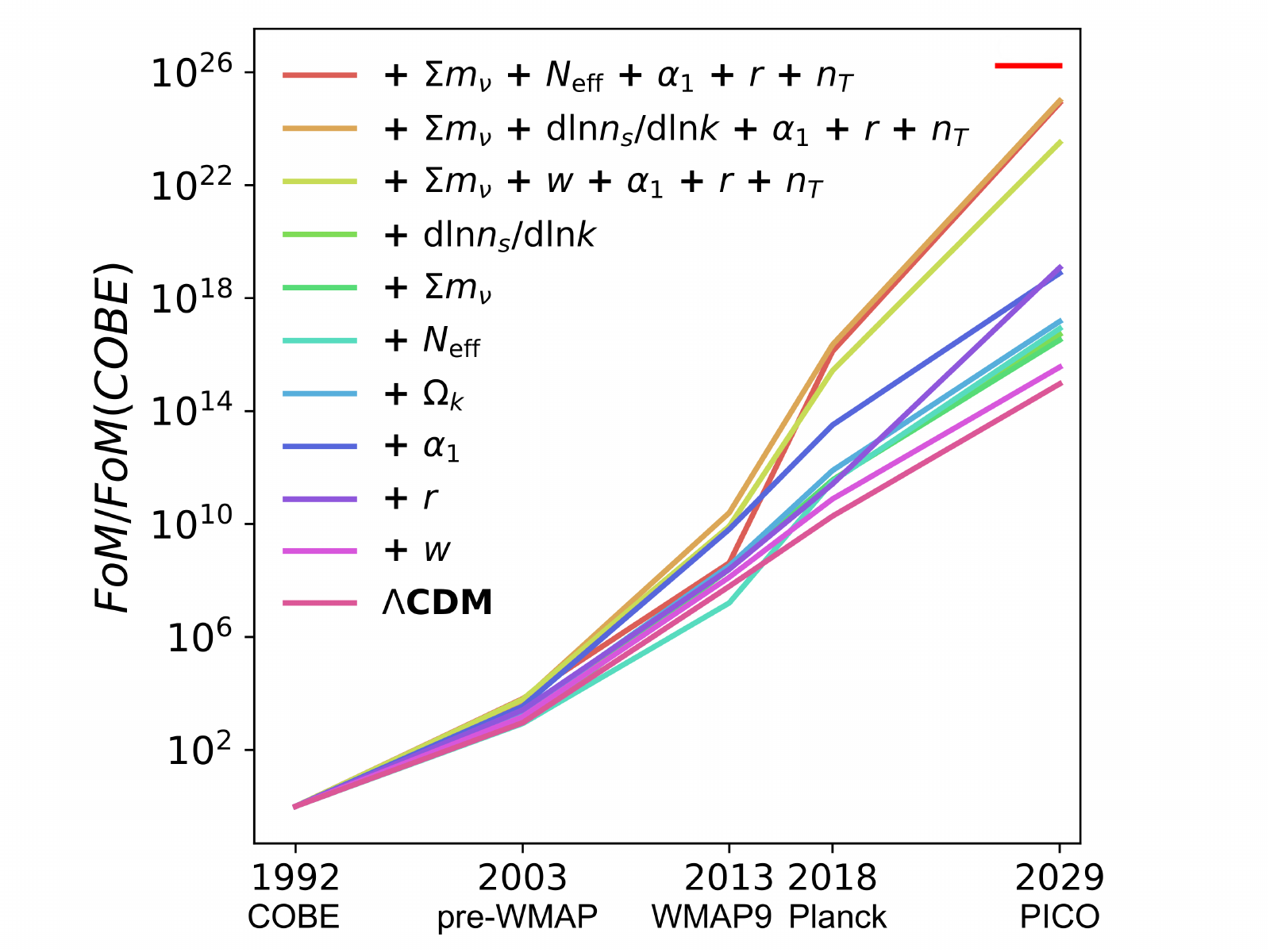} } }
\hspace{-0.17in}
\parbox{4.1in}{
\caption{\captiontext 
The increase in cosmological-parameter constraining power using only CMB data since \cobe . The FoM is the inverse of the uncertainly volume in parameter space. 
%for the $\Lambda$CDM six-parameter model (dark purple) and when adding other cosmological parameters. The FoM Increase in value represents increase in information content. PICO data will continue the average trend (blue line, $\Lambda$CDM + $\alpha_{1}$) of doubling the FOM every 10 months since 1992. 
For an 11-parameter set that includes $\Neff$ (red increasing line) PICO will improve the FoM by a factor of $5\times10^{8}$ relative to \planck . It will extract nearly the same information as that attainable by a mission with twice higher resolution and nine times lower noise (top right red horizontal bar), that is, PICO's performance on cosmological parameters is equivalent to that of a ``CMB flagship-scale mission''. The constituents of the 11-parameter set are given by~\citet{pico_report}. 
%includes: $w$-dark energy; $r$-the tensor to scalar ratio; $\alpha_{1}$-amplitude of correlated CDM isocurvature perturbations; $\Omega_{k}$-curvature; $\Neff$-effective number of light relics; $\sum m_{\nu}$-sum of neutrino masses; and $d\ln n_{s}/d\ln k$-running of the spectral index.   
\label{fig:fom} } }
\vspace{-0.1in}
\end{figure}

PICO will set the cosmological paradigm for the 2030's and beyond by measuring the six-parameter $\Lambda$CDM with 50,000 times more constraining power compared to \planck\ (Figure~\ref{fig:fom}; the improvement between \wmap\ and \planck\ was a factor of 300). For an 11-parameter set that includes $r$, $\Neff$, and $\Sigma (m_{\nu})$, the improvement is by a factor of $6\times10^{8}$. These improvements will test $\Lambda$CDM stringently. If it survives such scrutiny its dominance as the prevailing paradigm will strengthen. If tensions deepen to become discrepancies, it would be exciting to have a new cosmological model emerge. 

%The current host of cosmological observations including the CMB fit within the $\Lambda$CDM model, but they have statistical strength to constrain only six of more than a dozen known cosmological parameters. A figure of merit (FOM) that quantifies the strength of the constraint is the volume of the uncertainty region in the $N$-dimensional parameter space~\citep{core_parameter,Wang2008,pdg2018,Namikawa2010}. 

%Fig.~\ref{fig:fom} shows the increase in the FOM since \cobe\ for the six-parameter $\Lambda$CDM model, as well as for additional cosmological parameters. The Figure only includes data from CMB experiments. The FOM for $\Lambda$CDM improved by a factor of 100 between {\it WMAP} and \planck , and will further improve by a factor of $10^{5}$ with PICO. For the 11-parameter set that includes $\Neff$ shown in the Figure PICO will improve upon \planck\ by a factor of $0.5\times10^{9}$. Having achieved this improvement, there would be only little information left to extract with this parameter set even by a mission with double the resolution and nearly ten times lower noise (Fig.~\ref{fig:fom}). Even stronger FOM improvements are obtained when a 12-parameter set is considered~\citep{picoweb_lcdm}, and when the PICO CMB data will be combined with data sets available in the next decade, including weak lensing, BAO, and cluster of galaxies. 

%\end{document}

%% file: galacticsci-apc.tex
%\documentclass[PICOAPC.tex]{subfiles}

%\begin{document}

PICO will produce 21 polarization maps of Galactic emission, all much deeper than \planck 's seven maps. At 799~GHz PICO will have five times finer resolution than \planck . %; see Fig.~\ref{fig:allsky}). 
Such a dataset can only be obtained by a space mission like PICO. These data will complement a rich array of other polarization observations forthcoming in the next decade, including stellar polarization surveys to be combined with \gaia~astrometry, and Faraday rotation measurements from  observations at radio wavelengths with the  Square Kilometer Array and its precursors. \\
$\bullet$ {\bf Test Models of the Composition of Interstellar Dust} \hspace{0.1in}  
Less than a few $\mu$m in size, dust grains are intermediate in the evolution from atoms and molecules to large solid bodies such as comets, asteroids, and planets. Through vastly improved spectral characterization of Galactic polarization, the PICO data will 
%discriminate among models of Galactic dust composition to elucidate the chemical evolution of the Galaxy. For example, for the two-component paradigm \citep{Meisner2015}, the PICO baseline mission will determine the intrinsic polarization fractions of each of the two components to a precision of 3\%. With this level of precision the data will 
validate or reject state-of-the-art dust models~\citep[e.g.][]{Draine2009,Guillet2018,hensely_swp}, test for the presence of additional dust grain species with distinct polarization signatures, such as magnetic nanoparticles~\citep{Draine2013}, and will be used as an input for the foreground separation necessary to extract cosmological $E$- and $B$-mode science. \\
$\bullet$ {\bf Determine How Magnetic Fields Affect Molecular Cloud and Star Formation} \hspace{0.1in}
Stars are formed through interactions between gravitational and magnetic fields, turbulence, and gas over more than four orders of magnitude of spatial scale, which span the diffuse interstellar medium (ISM), molecular clouds, and molecular cloud cores. However, the role magnetic fields play in the large-scale structure of the ISM and in the observed low star-formation efficiency has been elusive, owing to the dearth of data. 
With 1.1~arcmin resolution, PICO will expand the number of independent magnetic field measurements across the sky from \planck 's  30,000 to 86,000,000, a factor of 2900. The data will robustly characterize turbulent properties like the Alfv\'{e}n Mach number across a previously unexplored regime of parameter space. 
%With full-sky coverage, PICO will map all the molecular clouds out to a distance of 3.4\,kpc with better than 1\,pc resolution; we estimate there will be over 2,000 clouds that will be mapped with $10^3$--$10^5$  independent polarization measurements per cloud. These are the {\it only foreseeable} measurements that will give the ratio of the energies stored in magnetic and gravitational fields, and the ratio of the energy stored in the magnetic field to that stored in gas turbulence over a statistically significant sample of molecular clouds.

%\end{document}

%\begin{figure}[!htb]
%\centering
%\includegraphics[width=4cm]{images/example}
%\caption{example}
%\label{fig:im_3}
%\end{figure}

%% file: Legacy-apc.tex
%\documentclass[PICOAPC.tex]{subfiles}

%\begin{document}

\definecolor{mygray}{gray}{0.6}

PICO will generate a rich and unique catalog of hundreds of thousands of new sources serving astrophysicists across a broad range of interests, including in galaxy and cluster evolution, correlations of cold Galactic dust with galactic properties, the physics of jets in active galactic nuclei, and the properties of the cosmic infrared background. This information will be embedded in catalogs including 50,000 proto-clusters extending to $z\simeq 4.5$, 4,500 strongly lensed galaxies extending to $z\simeq 5$, 30,000 galaxies with $z\leq 0.1$, polarization data for few thousand radio sources and dusty galaxies, and the deepest maps of the CIB with resolution as high as 1 arcmin. 

%\end{document}

%%%%%%%%%%%%%%%%%%%%%%%%%%%%%%%

%% file: foregrounds-systematics-apc.tex
%\documentclass[PICOAPC.tex]{subfiles}

%\newcolumntype{L}[1]{>{\raggedright\let\newline\\\arraybackslash\hspace{0pt}}m{#1}}
%\newcolumntype{K}[1]{>{\raggedright\centering\arraybackslash}m{#1}}

%\begin{document}

Controlling foregrounds and systematic effects are key for the success of any experimental endeavor striving to achieve $\sigma(r) \lesssim 1 \times 10^{-3}$. \\
$\bullet$ \hspace{0.1in}  PICO has the highest sensitivity of any next-decade CMB experiment, and the most frequency bands compared to any imaging instrument. It is thus more suitably equipped to handle foreground complexities. Higher sensitivity will translate to higher \ac{SNR} in detecting systematic effects. \\ 
$\bullet$ \hspace{0.1in}  \citet{pico_report} have shown that frequencies above 400~GHz may be essential for removing large-angular-scale foregrounds (see also \citet{hensley_2017}). They have also shown that for several realistic sky models PICO should be able to satisfy its $r$ detection requirement.  \\
%Based on community experience with both hardware and analysis of data we make the following points.  \\
$\bullet$ \hspace{0.1in}  Relative to other platforms, a space-based mission provides the most thermally stable platform,  a prerequisite for improved control of systematic effects. PICO's orbit at L2 is among the most thermally stable of possible orbits. \\
$\bullet$ \hspace{0.1in} PICO's sky scan pattern gives strong data redundancy, which enables numerous cross-checks. Each of the 12,996 detectors makes independent maps of the $I,\,Q$, and $U$ Stokes parameters, enabling many comparisons within and across frequency bands, within and across sections of the focal plane, and within and across bolometers that have either the same or different polarization sensitivities. Half the sky is scanned every two weeks, and the entire sky is scanned in 6 months. \\
$\bullet$ \hspace{0.1in}  The scan pattern gives almost continuous observations of planets and large amplitude ($\geq 4$~mK) CMB dipole signals~\citep{picoweb_dipole}. These features result in continuous, high \ac{SNR} calibration and antenna-pattern characterization. \\
We direct the reader to the mission study report for more details on foreground rejection and on characterization of systematic effects for PICO~\citep{pico_report}.

%\end{document}

%% file: technical-apc.tex
\newcommand\pdeg{.\!\!\degree}
\newcommand\parcm{.\!\!'}

\section{Technical Overview}
\label{sec:techoverview}

\vspace{-0.06in}

\begin{figure}[h] 
\hspace{-0.1in}
\parbox{4.8in}{\centerline{
\includegraphics[width=4.8in]{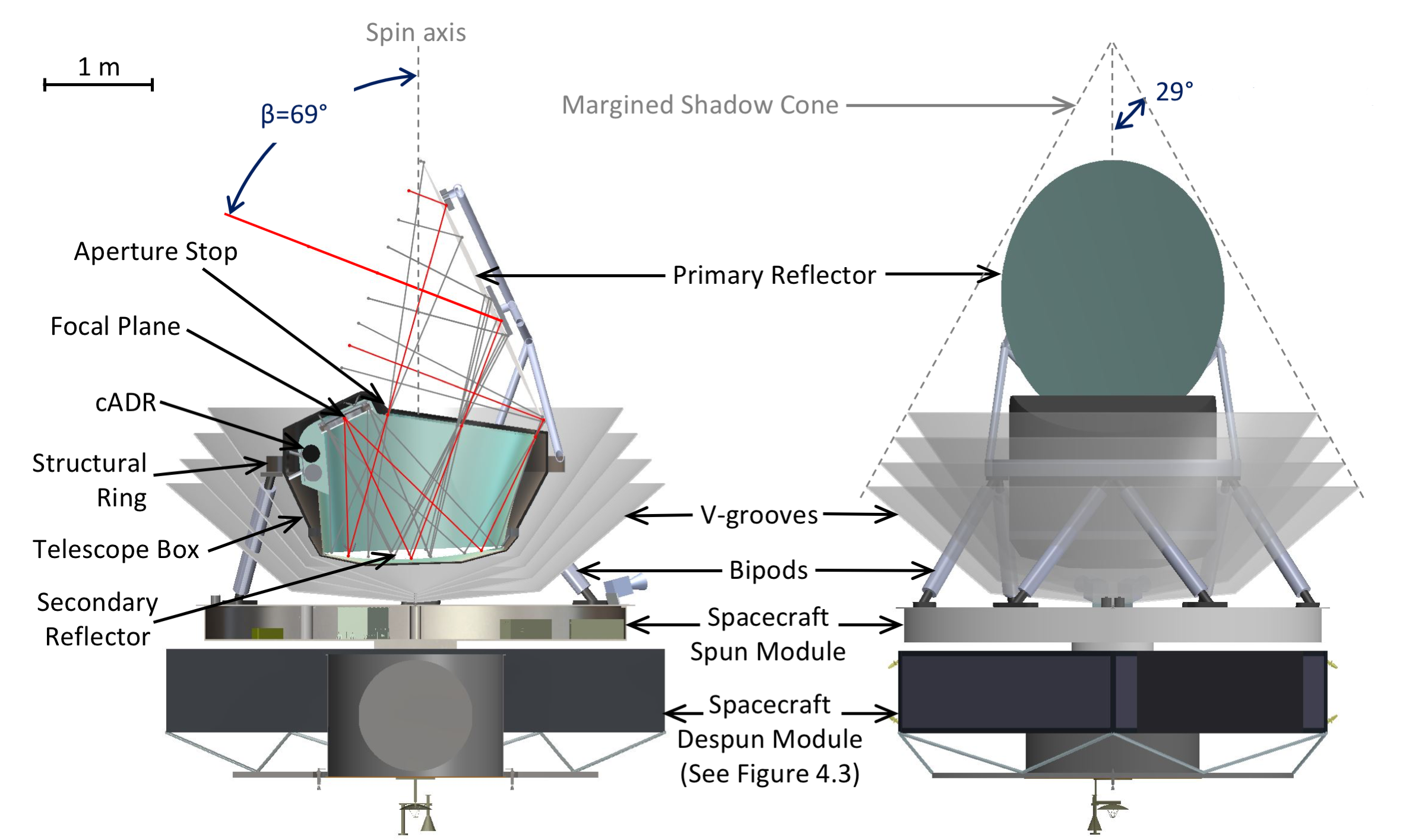} }}
\hspace{0.0in}
\parbox{1.6in}{
\caption{\captiontext
PICO overall configuration in side view and cross section (left), and front view with V-Groove assembly shown semi-transparent (right).  The mission consists of a single science instrument mounted on a structural ring. The ring is supported by bipods on a stage spinning at 1~RPM relative to a despun module. Only power and digital information pass between the spun and despun stages. 
\label{fig:InstrumentCAD}} }
%\end{center}
\vspace{-0.15in}
\end{figure}

PICO meets all of its science-driven instrument requirements with a single instrument: an imaging polarimeter with 21 logarithmically spaced frequency bands centered between 21 and 799\,GHz (Table~\ref{tab:spec_bands}). The instrument has a two-reflector Dragone-style telescope; see Figure~\ref{fig:InstrumentCAD}. The focal plane is populated by \ac{TES} bolometers and read out using a time-domain multiplexing scheme. The instrument has both passive and active cooling stages. PICO operates from the Earth-Sun L2 and employs a single science observing mode, providing highly redundant coverage of the full sky. A full description of the reference design is given by \citet{pico_report}.

\vspace{-0.06in}

\subsection{Telescope, Detectors, and Readout}
\label{sec:telescope} 

\vspace{-0.03in}

The PICO 1.4~m-aperture, two-mirror telescope gives: a large diffraction-limited field of view, sufficient to support approximately $10^4$ detectors; arcminute resolution at 800\,GHz; low instrumental and cross-polarization; and low sidelobe response. There are no moving parts in the PICO optical system, reducing mission risk. There are no lenses, eliminating absorption and reflection losses and obviating the need for developing broad-band anti-reflection coatings. The primary mirror is passively cooled to $\sim$20~K. An aperture stop and a secondary mirror are actively cooled to 4.5~K. 

%The sensitivity of PICO's detectors is limited by irreducible backgrounds: the CMB at frequencies below 400~GHz, and emission from the cold primary mirror at higher frequencies. Therefore, the required sensitivity determines the detector count in each band. 
The PICO focal plane has a total of 12,996 \ac{TES} detectors, 175~times the number flown aboard \planck . The required full-sky, 5-year survey depth is 0.87~$\mu$K\,arcmin; the current best estimate performance is 0.61\,$\mu$K\,arcmin, offering 40\% noise margin. PICO is the most sensitive CMB experiment proposed for the next decade. To achieve similar raw sensitivity, a ground-based instrument would require $\simgt$50 times the number of detectors. 

\input{table-bands.tex}

There is broad flexibility in the detailed implementation of the PICO focal plane. In the baseline design we employ three-color sinuous antenna/lenslet pixels~\citep{Suzuki2014} for the 21--462\,GHz bands, and single color, feedhorn-coupled, polarization sensitive bolometers for the three higher frequency bands. PICO can also achieve its required performance with two-color pixels~\citep{nist_design} (for 21--462~GHz) with a total of 19 bands and the same noise margin, or even with single-color pixels at all 21 bands~\citep{bicep_design} and 17\% noise margin. Current ground-based instruments use all three technologies at a narrower range of frequencies. There are funding mechanisms and development programs in place to adapt the technologies to a broader range of frequencies and to space applications; see Section~\ref{sec:techdrivers}. 
Recent results from balloon experiments indicate that modern bolometer arrays are less susceptible to cosmic-ray energy deposition compared to the individual detector design used with \planck . We recommend a program of testing and characterization; see Section~\ref{sec:techdrivers}. 

%PICO can achieve its required performance with other already focal plane technologies  than albeit with reduce
%Niobium microstrips mediate the signals between the antenna and detectors, and partition the wide continuous bandwidth into three narrow channels using integrated, on-wafer, micro-machined filter circuits~\citep{OBrient2013}. Six transition edge sensor bolometers per pixel detect the radiation in two orthogonal polarization states. 

%PICO's highest three frequency channels are beyond the niobium superconducting band-gap, rendering on-wafer, microstrip filters a poor solution for defining the optical passband. For these bands we use feedhorns to couple the radiation to two single-color polarization-sensitive TES bolometers. The waveguide cut-off defines the lower edge of the band, and quasi-optical metal-mesh filters define the upper edge. Numerous experiments have successfully used similar approaches~\citep{Shirokoff2011,Bleem2012,Turner2001}. 

Polarimetry is achieved by measuring the signals from pairs of two co-pointed bolometers within a pixel that are sensitive to two orthogonal linear polarization states. Half the pixels in the focal plane are sensitive to the $Q$ and half to the $U$ Stokes parameters of the incident radiation, providing sensitivity to the Stokes $I$, $Q$, and $U$ parameters. Two layouts for the distribution of the $Q$ and $U$ pixels on the focal plane have been investigated~\citep{picoweb_QU}; both satisfy mission requirements. 

The current baseline for PICO is to use a time-domain multiplexer (TDM), because to date this scheme uses the least power consumption and dissipation at ambient temperatures. The thermal loading on the cold stages from the wire harnesses is subdominant to other loadings. In the PICO TDM implementation, a row of 102 detectors are read out simultaneously, and 128 such rows are read out sequentially. SQUIDs will be used as current amplifiers. All the technology elements necessary for implementing this readout have already been demonstrated~\citep{pico_report}. Only packaging for space is required; see Section~\ref{sec:techdrivers}. Suborbital experiments have developed techniques to shield the SQUIDs from Earth's magnetic field~\citep{Hui2018}.  PICO will use these techniques to shield SQUID readout chips from the ambient magnetic environment, which is 20,000 times weaker than near-Earth. 

\vspace{-0.06in}

\subsection{Thermal}
\label{sec:thermal} 

\vspace{-0.03in}

The PICO thermal system does not require cryogenic consumables, permitting consideration of significant mission extension beyond the prime mission. The system, consisting of V-groove radiators for passive cooling, mechanical coolers to achieve 4.5\,K, and a continuous adiabatic demagnetization refrigerator (cADR), meets all thermal requirements with robust margins~\citep{pico_report} .

The cADR maintains the focal plane at 0.1\,K and the surrounding enclosure, filters, and readout components at 1\,K. Heat loads in the range of 30\,$\mu$W at 0.1\,K and 1\,mW at 1\,K (time-average) are within the capabilities of current cADRs developed by GSFC~\citep{Shirron2012,Shirron2016} and flown on suborbital balloon flights. The PICO sub-kelvin heat loads are estimated at less than half of this capability.

A cryocooler system similar to that used on \jwst~to cool the MIRI detectors~\citep{Durand2008,Rabb2013} backs the cADR and cools the aperture stop and secondary reflector to 4.5\,K. Both Northrop Grumman Aerospace Systems (NGAS, which provided the MIRI coolers) and Ball Aerospace have developed such coolers under the NASA-sponsored Advanced Cryocooler Technology Development Program~\citep{Glaister2006}. The NGAS project has completed PDR-level development, and is expected to reach CDR well before PICO begins Phase-A. The projected performance of this cooler will give more than 100\,\% heat lift margin relative to PICO's requirements~\citep{pico_report}.

%\vspace{-0.06in}

%\subsection{Instrument Integration and Test}
%\label{sec:iandt} % 3.5

%\vspace{-0.03in}

%The PICO instrument integration and testing plan benefits from heritage and experience with the \planck\ HFI instrument~\citep{Pajot2010}.
%Detector wafers are screened prior to selection of flight wafers and focal-plane integration. The cADR and 4\,K cryocooler vendors will qualify those subsystems prior to delivery. The relative alignment of the two reflectors is determined under in-flight thermal conditions using a thermal vacuum (TVAC) chamber and photogrammetry. The flight focal-plane assembly and flight cADR are integrated and tested in a dedicated sub-kelvin cryogenic testbed. The noise, responsivity, and focal-plane temperature stability are characterized using a representative optical load for each frequency band (temperature-controlled blackbody).  The same testbed is used to perform the  polarimetric and spectroscopic calibration.

%The focal plane is integrated with the reflectors and structures, and alignment verified with photogrammetry at cold temperatures in a TVAC chamber.  The completely integrated observatory (instrument and spacecraft bus) is tested in TVAC to measure parasitic optical loading from the instrument, noise, microphonics, and RFI. The observatory is 4.5\,m in diameter and 6.1\,m tall. There are no deployables.

\vspace{-0.06in}

\subsection{Design Reference Mission}
\label{sec:design_reference} 

\vspace{-0.03in}

\begin{figure}[h] 
\hspace{-0.5in}
\parbox{4.0in}{\centerline{
\includegraphics[width=2.5in]{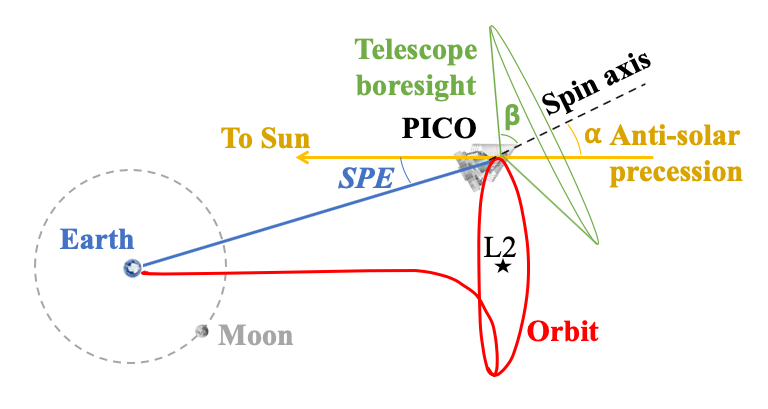} } }
\hspace{-0.6in}
\parbox{3.5in}{
\caption{\captiontext
PICO surveys the sky by spinning the instrument about the spacecraft's symmetry axis at 1~RPM. The telescope boresight is tilted by $\beta=69^{\rm{o}}$ from that axis. The symmetry axis precesses around the anti-Sun direction with a period of 10~hours; $\alpha=29^{\rm{o}}$. Nearly 50\% of the sky is surveyed every two weeks. The entire sky is covered in 6 months. 
\label{fig:MissionDesignFigure}} }
%\end{center}
\vspace{-0.15in}
\end{figure}

The PICO concept of operations is similar to that of the successful \wmap~\citep{Bennett2003} and \planck~\citep{Tauber2010} missions. After launch, PICO cruises to a quasi-halo orbit around the Earth--Sun L2 Lagrange point. A two-week decontamination period is followed by instrument cooldown, lasting about two months. After in-orbit checkout is complete, PICO begins its science survey, depicted in Figure~\ref{fig:MissionDesignFigure}.  This survey ensures that each sky pixel is revisited along many orientations, which is optimal for polarimetric measurements. 
%Nearly 50\% of the sky are surveyed every two weeks. The entire sky is covered in 6 months. 
Over the 5-year duration, PICO executes 10 independent full-sky surveys, giving high redundancy for identifying systematic uncertainties. 

Instrument data are compressed and stored on-board, then returned to Earth in daily 4-hr Ka-band science downlink passes. High data-rate downlink to the Deep Space Network (DSN) is available from L2 using near-Earth Ka bands. We have assumed a launch with the Falcon~9; its capability for ocean recovery exceeds PICO's 2147-kg total launch mass (including contingency) by a $50\,\%$ margin.

The PICO spacecraft bus is Class~B and is designed for a minimum lifetime of 5\,years in the L2 environment. Mission-critical elements are redundant. The aft end of the spacecraft (the ``de-spun module'') is comprised of six equipment bays that house standard components.  The instrument and V-grooves are mounted on bipods from the spacecraft's ``spun module,'' which contains the 4-K cooler compressor and drive electronics, the sub-K cooler drive electronics, and the detector warm readout electronics. A motor drives the spun module at 1\,rpm. Only power and digital data lines pass between the spun and de-spun modules. Reaction wheels on the despun module cancel the angular momentum of the spun module and provide three-axis control.

%% file: table-bands.tex
%\begin{wraptable}[13]{r}{0.43\textwidth}
%\vskip -2mm
%\hskip1.4cm
%\begin{minipage}[t]{0.3\textwidth}
%\caption{\textbf{Mission Parameters}\label{tab:specs}}
%\begingroup
%%\openup 5pt
%\newdimen\tblskip \tblskip=5pt
%\nointerlineskip
%\vskip -7mm
%\footnotesize %\footnotesize
%\setbox\tablebox=\vbox{
%    \newdimen\digitwidth
%    \setbox0=\hbox{\rm 0}
%    \digitwidth=\wd0
%    \catcode`*=\active
%    \def*{\kern\digitwidth}
%%
%    \newdimen\signwidth
%    \setbox0=\hbox{+}
%    \signwidth=\wd0
%    \catcode`!=\active
%    \def!{\kern\signwidth}
%%
%\halign{%
%\hbox to 1.4in{#\leaderfil}\tabskip=0.6em plus 0.6em&
%#\hfil\tabskip=0pt\cr
%\noalign{\doubleline}
%\multispan2 Full sky CMB polarization map depth$^{a}$ :\hfil\cr
%\quad Baseline&0.87 $\mu$K$_{\rm CMB}$ arcmin\cr
%\hskip1.75cm equivalent to 3300 \textit{Planck} missions\cr
%\quad CBE$^{b}$&0.61 $\mu$K$_{\rm CMB}$ arcmin\cr
%\hskip1.75cm equivalent to 6400 \textit{Planck} missions\cr
%Survey duration / start & 5\,yrs / 2029 \cr
%Orbit type & Sun-Earth L2 \cr
%Launch mass & 2147\,kg \cr
%Total power &1320\,W \cr
%Data rate & 6.1\,Tbits/day \cr
%Cost&\$\,958M\cr
%%Launch&2029\cr
%\noalign{\vskip 5pt\hrule\vskip 3pt}
%\noalign{$^{a}$ rms noise in $1\times1$ arcmin$^{2}$ pixel.} % footnote to table
%\noalign{$^{b}$ CBE = Current best estimate.} % footnote to table
%} % close halign
%} % close vbox
%\endPlancktable
%\endgroup
%%\end{table}
%\end{minipage}

%\end{wraptable}

% Start second table
%\vskip0.5cm
%\hskip0.2cm

\begin{wraptable}[24]{r}{0.43\textwidth}
\vskip -2mm
%\hskip1.4cm
\begin{minipage}[t]{0.4\textwidth}
\footnotesize
\caption{\textbf{Frequency Bands, Resolution, and Noise Level}\label{tab:spec_bands}}
\vskip -2mm
\begin{tabular}{|c|c|c|c|}
\hline
Frequency& FWHM& \multicolumn{2}{c|}{Polarization map depth}  \\ 
\cline{3-4}
& & Baseline & CBE  \\
 \space [GHz] & [arcmin]  & [$\mu$K$_{\rm CMB}$]$^{a}$ & [$\mu$K$_{\rm CMB}]^{a}$ \\ \hline
21 & 38.4 & 23.9 & 16.9 \\ 
25 & 32.0 & 18.4 & 13.0 \\ 
30& 28.3 & 12.4 & 8.7 \\ 
36& 23.6 & 7.9 & 5.6 \\ 
43& 22.2 & 7.9 & 5.6 \\ 
52& 18.4 & 5.7 & 4.0 \\ 
62& 12.8 & 5.4 & 3.8 \\ 
75& 10.7 & 4.2 & 3.0 \\ 
90 & 9.5 & 2.8 & 2.0 \\ 
108 & 7.9 & 2.3 & 1.6\\ 
129 & 7.4 & 2.1 & 1.5 \\ 
155 & 6.2 & 1.8 & 1.3 \\ 
186 & 4.3 & 4.0 & 2.8 \\ 
223 & 3.6 & 4.5 & 3.2 \\ 
268 & 3.2 & 3.1 & 2.2 \\ 
321 & 2.6 & 4.2 & 3.0 \\ 
385 & 2.5 & 4.5 & 3.2 \\ 
462 & 2.1 & 9.1 & 6.4 \\ 
555 & 1.5 & 45.8 & 32.4 \\ 
666 & 1.3 & 177 & 125 \\ 
799 & 1.1 & 1050 & 740 \\ 
\noalign{\vskip 0pt\hrule\vskip 3pt}
\noalign{$^{a}$For values in [Jy/sr] see~\citet{pico_report}.} % footnote to table

\end{tabular}

%\end{table}
\end{minipage}
\end{wraptable}

%% file: tech-drivers-apc.tex
The remaining technology developments required to enable the PICO baseline design are: \\
$\bullet$ \hspace{0.05in} extension of three-color antenna-coupled bolometers down to 21\,GHz and up to 462\,GHz; \\
$\bullet$ \hspace{0.05in} construction of high-frequency direct absorbing arrays and laboratory testing; \\
$\bullet$ \hspace{0.05in} beam line and 100-mK testing to simulate the cosmic ray environment at L2; and \\
$\bullet$ \hspace{0.05in} implementation of time-domain multiplexing to support 128 switched rows per readout column. \\
All of these developments are straightforward extensions of technologies already available and used today by sub-orbital and orbital experiments.  We recommend APRA and SAT funding to complete the development through the milestones described in Table~\ref{tab:technologies}~\citep{pico_report}.

\input table-tech-drivers.tex

%% file: table-tech-drivers.tex
%% Table 5.1 in the PICO report

\begin{table}
\caption{\captiontext 
  PICO technologies can be developed to TRL~5 prior to a 2023 Phase~A start using the APRA ("A") and SAT ("S") programs, requiring an estimated total of \$13M. \label{tab:technologies}}
\begingroup
%\openup 5pt
\newdimen\tblskip \tblskip=5pt
\nointerlineskip
\vskip -7mm
\footnotesize %\footnotesize
\setbox\tablebox=\vbox{
    \newdimen\digitwidth
    \setbox0=\hbox{\rm 0}
    \digitwidth=\wd0
    \catcode`*=\active
    \def*{\kern\digitwidth}
    \newdimen\signwidth
    \setbox0=\hbox{+}
    \signwidth=\wd0
    \catcode`!=\active
    \def!{\kern\signwidth}
\halign{
\vtop{\hsize 1.3in\raggedright\hangafter=1\hangindent=2em\noindent\strut#\strut\par}\leaderfil\tabskip=0.6em&
\vtop{\hsize 0.7in\raggedright\hangafter=1\hangindent=0em\noindent\strut#\strut\par}&
\vtop{\hsize 0.9in\raggedright\hangafter=1\hangindent=0em\noindent\strut#\strut\par}&
\vtop{\hsize 0.7in\raggedright\hangafter=1\hangindent=0em\noindent\strut#\strut\par}&
\vtop{\hsize 0.6in\raggedright\hangafter=1\hangindent=0em\noindent\strut#\strut\par}&
\vtop{\hsize 0.5in\raggedright\hangafter=1\hangindent=0em\noindent\strut#\strut\par}&
\vtop{\hsize 0.7in\raggedright\hangafter=1\hangindent=0em\noindent\strut#\strut\par}&
\hfil#\hfil \tabskip=0pt\cr
\noalign{\doubleline}
\omit\hfil Task\hfil&\omit\hfil Current\hfil&\omit\hfil Milestone A\hfil&\omit\hfil Milestone B\hfil&\omit\hfil Milestone C\hfil&\omit\hfil Current\hfil&\omit\hfil Required\hfil&\omit\hfil Date TRL5\hfil\cr
\omit&\omit\hfil status\hfil&&&&\omit\hfil funding\hfil&\omit\hfil funding\hfil&\omit\hfil achieved\hfil\cr
\noalign{\vskip 3pt\hrule\vskip 5pt}
1a. Three-color arrays $\nu<90$\,GHz&2-color lab demos $\nu > 30$\,GHz&Field demo of 30--40\,GHz (2020)&Lab demos 20--90 GHz (2022)& \hspace{0.12in} ----- &A \& S&\$2.5M over 4\,yr (1 A + 1 S) &2022\cr
1b. Three-color arrays $\nu > 220$\,GHz&2-color lab demos $\nu < 300$\,GHz&Field demo of 150--270\,GHz (2021)&Lab demos 150-460 GHz (2022) &\hspace{0.12in} ----- &A \& S&\$3.5M over 4\,yr (2 Ss) & 2022\cr
2. Direct absorbing arrays $\nu > 550$\,GHz& 0.1--5\,THz unpolarized&Design \& prototype of arrays (2021)&Lab demo of 555\,GHz (2022)& Lab demo 799\,GHz (2023) & None & \$2M over 5\,yr (1 S) & 2023 \cr
3. Cosmic ray studies& 250\,mK w/ sources&100\,mK tests with sources (2021)&Beamline tests (2023)& \hspace{0.12in} ----- & A \& S & \$0.5--1M over 5\,yr (part of 1 S) & \hspace{0.0in} ----- \cr
4a. Fast readout electronics& MUX66 demo& Electronics Engineering and Fab (2020)&Lab demo (2021)&Field demo (2023)& No direct funds &\$4M over 5\,yr (1 S)&2023\cr
4b. System engin\-eering; 128$\times$ MUX demo&MUX66 demo&Design of cables (2020)&Lab demo (2021)& Field demo (2023) & No direct funds & \hspace{0.12in} ----- & \hspace{0.0in} ----- \cr
\noalign{\vskip 5pt\hrule\vskip 3pt}
} % close halign
} % close vbox
\endPlancktable
\endgroup
\vskip -3mm
\end{table}

%% file: organization-apc.tex
\section{Organization, Partnerships, and Current Status}
\label{sec:organization} %6

%\subsection{PICO Study Participants}
%\label{sec:study_participants} %6.1

PICO is the result of an 18-month mission study funded by NASA (total grant = \$150,000). The study was open to the entire mm/sub-mm community. Seven working groups were led by members of PICO's Executive Committee, which had a telephone conference weekly, led by the PI. A three-member steering committee, composed of two experimentalists experienced with CMB space missions, and a senior theorist gave occasional advice to the PI. More than 60 scientists, international- and US-based, participated in-person in each of two community workshops (November 2017 and May 2018). The study report has been submitted by NASA to the decadal panel, and it is available on the arXiv and on the PICO website~\cite{pico_report, picoweb}. It has contributions from 82 authors, and has been endorsed by an additional 131 members of the community. 

%The PICO engineering concept definition package including the total cost was generated by Team~X.\footnote{\label{teamx} Team~X is JPL's concurrent design facility.} The Team~X study was supported by inputs from a JPL engineering team and Lockheed Martin.

The PICO team designed an entirely US-based mission, so that the full cost of the mission can be assessed. We excluded contributions by other space agencies, despite expression of interest by international scientists. The PICO concept has wide support in the international community. If the mission is selected to proceed, 
%a path that would be scientifically and financially optimal relative to other options, 
it is reasonable to expect that international partners would participate and thus reduce the US cost of the mission.

%\end{document}

%% file: schedule-and-cost-apc.tex
\section{Schedule and Cost}
\label{sec:mission_schedule}

%\subsection{Schedule}
%\label{sec:mission_schedule}

\vspace{-0.1in}
\begin{figure}[h]
\begin{center}
\includegraphics[width=\textwidth]{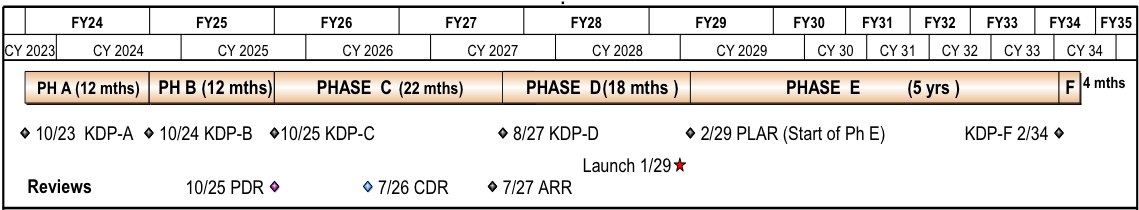}
\end{center}
\vspace{-0.30in}
\caption{\captiontext PICO development and operations schedule. \label{fig:Schedule}}
\vspace{-0.05in}
\end{figure}

$\bullet$ {\bf Schedule} \hspace{0.1in} NASA-funded Probe studies including PICO assume a Phase~A start in October 2023. PICO development phases B--D are similar in duration to recent comparably sized NASA missions such as Juno and SMAP. PICO is a cryogenic mission similar to \planck , but the cryogenic design is simpler because all of PICO's detectors are maintained at 0.1~K (some of \planck 's detectors were maintained at 0.1\,K, and some at 20~K).  We used experience from \planck\ and from current implementations of ground-based kilo-pixel arrays  to allocate appropriate time for integration and testing. The baseline mission lifetime is 5 years. 
%The PICO instrument does not have cryogenic consumables (as \planck\ did) permitting mission extension beyond the prime mission duration. 

%\newpage
\input table-cost-abridged.tex
\noindent $\bullet$ {\bf Cost} \hspace{0.1in}  We estimate PICO's total Phase A--E lifecycle cost between \$870M and \$960M, including the \$150M allocation for the Launch Vehicle (per NASA direction). These cost estimates include reserves; see Table~\ref{tab:cost}. The Table shows the 
JPL Team~X and the PICO team mission cost breakdown. Team~X estimates are generally model-based, and were generated after a series of instrument and mission-level studies. The PICO team adopted the Team~X estimates, but also obtained a parametrically estimated cost range for the spacecraft bus, assembly, test, and launch operations from Lockheed Martin Corporation to represent the cost benefits that might be realized by working with an industry partner. After adding estimated JPL overhead 
%and Team~X estimated V-groove assembly costs (not included in the Lockheed estimate), 
the PICO team cost is in-family with but lower than the Team~X cost.

%Science team costs are assessed by Team~X based on PICO science team estimates of the numbers and types of contributors and meetings required for each year of PICO mission development and operations. These workforce estimates are informed by recent experience with the \planck\ mission. 

PICO's spacecraft cost reflects a robust Class~B architecture. Mission-critical elements are redundant. Appropriate flight spares, engineering models and prototypes are included. Mission operations, ground data systems, and mission navigation and design costs reflect the relatively simple operations: PICO has a single instrument and a single, repetitive science observing mode. The mission costs include \$19M for the science team during development and \$20M during operations, based on science team workforce estimates informed by recent experience with the \textit{Planck} mission. Pre-Phase-A technology maturation will be accomplished through the normal APRA and SAT processes, and is not included in the mission cost. The PICO Team estimates the total cost of the required technology maturation to be about \$13M (Table~\ref{tab:technologies}).

%% file: table-cost-abridged.tex
%% Table 6.1 abridged
\begin{wraptable}[11]{r}{0.52\textwidth}
\vspace{-7pt} \hspace{0.01\textwidth}
\begin{minipage}{0.48\textwidth}
\caption{\captiontext PICO mission costs}\label{tab:cost}
\vspace{-15pt}
\begingroup
%\openup 5pt
\newdimen\tblskip \tblskip=5pt
\nointerlineskip
\vskip -3mm
\footnotesize %\footnotesize
\setbox\tablebox=\vbox{
    \newdimen\digitwidth
    \setbox0=\hbox{\rm 0}
    \digitwidth=\wd0
    \catcode`*=\active
    \def*{\kern\digitwidth}
    \newdimen\signwidth
    \setbox0=\hbox{+}
    \signwidth=\wd0
    \catcode`!=\active
    \def!{\kern\signwidth}
\halign{
\hbox {\vtop{\hsize1.95in\hangafter=1\hangindent=2em\noindent\raggedright\strut#\strut\par}}\tabskip=0.6em&
\hfil#\hfil&
\hfil#\hfil\tabskip=0pt\cr
\noalign{\doubleline}
\omit&\multispan2 \hfil \textbf{Estimate by}\hfil\cr
\omit&JPL&PICO \cr
\hfil \textbf{Project Phase}\hfil&Team X&Team \cr
\noalign{\vskip 3pt\hrule\vskip 5pt}
Development (Phases A--D)&\$\,724M&\$\,634--677M\cr
\hspace{6pt} (includes 30\% reserves) &&\cr
Operations (Phases E--F)&\multispan2 \hfil \$\,*84M\hfil\cr
\hspace{6pt} (includes 13\% reserves) &&\cr
Launch Vehicle &\multispan2 \hfil\$\,150M\hfil\cr
\noalign{\vskip 3pt\hrule\vskip 5pt}
\textbf{Total Cost (FY18\$)}&\$\,958M&\$\,868--911M\cr
\noalign{\vskip 5pt\hrule\vskip 3pt}
} % close halign
} % close vbox
\endPlancktable
\endgroup
\end{minipage}
\end{wraptable}

%% file: acronym.tex
\begin{acronym}
    %A
    \acro{ACS}{attitude control system}
    \acro{ADC}{analog-to-digital converters}
    \acro{ADS}{attitude determination software}
    \acro{AHWP}{achromatic half-wave plate}
    \acro{AMC}{Advanced Motion Controls}
    \acro{AME}{anomalous microwave emission}
    \acro{ARC}{anti-reflection coatings}
    \acro{ATA}{advanced technology attachment}
    %B
    \acro{BAO}{baryon acoustic oscillations}
    \acro{BRC}{bolometer readout crates}
    \acro{BLAST}{Balloon-borne Large-Aperture Submillimeter Telescope}
    %C
    \acro{CANbus}{controller area network bus}
    \acro{CBE}{current best estimate}
    \acro{CIB}{cosmic infrared background}
    \acro{CMB}{cosmic microwave background}
    \acro{CMM}{coordinate measurement machine}
    \acro{CSBF}{Columbia Scientific Balloon Facility}
    \acro{CCD}{charge coupled device}
    %D
    \acro{DAC}{digital-to-analog converters}
    \acro{DASI}{Degree~Angular~Scale~Interferometer}
    \acro{dGPS}{differential global positioning system}
    \acro{DfMUX}{digital~frequency~domain~multiplexer}
    \acro{DLFOV}{diffraction limited field of view}
    \acro{DSP}{digital signal processing}
    %E
    \acro{EBEX}{E~and~B~Experiment}
    \acro{EBEX2013}{EBEX2013}
    \acro{ELIS}{EBEX low inductance striplines}
    \acro{ETC}{EBEX test cryostat}
    %F
    \acro{FDM}{frequency domain multiplexing}
    \acro{FPGA}{field programmable gate array}
    \acro{FCP}{flight control program}
    \acro{FOV}{field of view}
    \acro{FWHM}{full width half maximum}
    %G
    \acro{GPS}{global positioning system}
    %H
    \acro{HDPE}{high density polyethylene}
    \acro{HIM}{high index materials}
    \acro{HWP}{half-wave plate} 
    %I
    \acro{IA}{integrated attitude}
    \acro{ICM}{intercluster medium}
    \acro{IGM}{intergalactic medium}
    \acro{IGW}{inflationary gravity wave} 
    \acro{ILC}{independent linear combination}
    \acro{IP}{instrumental polarization} 
    \acro{ISM}{interstellar medium}
    %J
    \acro{JSON}{JavaScript Object Notation}
    %L
    \acro{LDB}{long duration balloon}
    \acro{LED}{light emitting diode}
    \acro{LC}{inductor and capacitor}
    \acro{LCS}{liquid cooling system}
    \acro{LZH}{Lazer Zentrum Hannover}
%M
    \acro{MCP}{multi-color pixel}
    \acro{MSM}{millimeter and sub-millimeter}    
    \acro{MLR}{multilayer reflective}
    \acro{MAXIMA}{Millimeter~Anisotropy~eXperiment~IMaging~Array}
    %N
    \acro{NASA}{National Aeronautics and Space Administration}
    \acro{NDF}{neutral density filter}
    %P
    \acro{PCB}{printed circuit board}
    \acro{PE}{polyethylene}
%    \acro{PTFE}{polytetrafluoroethylene}
    \acro{PME}{polarization modulation efficiency}
    \acro{PSF}{point spread function}
    \acro{PV}{pressure vessel}
    \acro{PWM}{pulse width modulation}
    %R
    \acro{RMS}{root mean square}
%S
    \acro{SED}{spectral energy distribution}
    \acro{SLR}{single layer reflective}
    \acro{SMB}{superconducting magnetic bearing}
    \acro{SNR}{signal-to-noise ratio}
    \acro{SOs}{science objectives}
    \acro{SO}{science objective}
    \acro{SQUID}{superconducting quantum interference device}
    \acro{SQL}{structured query language}
    \acro{STARS}{star tracking attitude reconstruction software}
    \acro{SZ}{Sunyaev--Zeldovich}
    \acro{SWS}{sub-wavelength structures}
%T
    \acro{tSZ}{thermal Sunyaev--Zeldovich}
    \acro{TES}{transition-edge-sensor}
    \acro{TDRSS}{tracking and data relay satellites}
   \acro{TM}{transformation matrix}
   \acro{TRL}{Technology Readiness Level}
% U
    \acro{UHMWPE}{ultra high molecular weight polyethylene}   
    \acro{UMN}{University of Minnesota}
    
\end{acronym}